\documentclass[journal,onecolumn]{IEEEtran}

\usepackage{cite}

\usepackage{amsmath,amssymb,amsfonts, amsthm}
\usepackage{graphicx}
\usepackage{textcomp}
\usepackage{dsfont}
\usepackage{algorithm}
\usepackage{algpseudocode}
\usepackage{booktabs}
\usepackage{siunitx}
\usepackage{xcolor}
\usepackage{booktabs}
\usepackage[framemethod=tikz]{mdframed}
\usetikzlibrary{calc}
\allowdisplaybreaks
\usepackage{bm}

\renewcommand{\P}[1]{\mathbb{P}\left[#1\right]}
\renewcommand{\vec}{\bm}

\newcommand{\M}{\mathcal{M}}
\newcommand{\E}[2][]{\mathbb{E}_{#1}\left[#2\right]}
\DeclareMathOperator*{\minimize}{minimize}
\newcommand{\norm}[2][2]{\left\Vert {#2} \right\Vert^{#1}}

\newtheorem{defi}{Definition}
\newtheorem{theo}[defi]{Theorem}

\DeclareMathOperator*{\argmin}{arg\,min}

\newcommand{\thickhline}{
    \noalign {\ifnum 0=`}\fi \hrule height 1pt
    \futurelet \reserved@a \@xhline
}

  \usepackage[caption=false,font=normalsize,labelfont=sf,textfont=sf]{subfig}

\tikzset{every picture/.style={line width=0.75pt}}

\begin{document}
\title{Massive Random Access with Common Alarm Messages}
\author{\IEEEauthorblockN{Kristoffer Stern, Anders E. Kal{\o}r, Beatriz Soret, Petar Popovski}\\
\IEEEauthorblockA{Department of Electronic Systems, Aalborg University, Denmark\\
Email: kstern14@student.aau.dk, \{aek, bsa, petarp\}@es.aau.dk}
}

\maketitle

\begin{abstract}
The established view on massive IoT access is that the IoT devices are activated randomly and independently. This is a basic premise also in the recent information-theoretic treatment of massive access by Polyanskiy \cite{Polyanskiy2017}. In a number of practical scenarios, the information from IoT devices in a given geographical area is inherently correlated due to a commonly observed physical phenomenon. We introduce a model for massive access that accounts for correlation both in device activation and in the message content. To this end, we introduce common alarm messages for all devices. A physical phenomenon can trigger an alarm causing a subset of devices to transmit the same message at the same time. We develop a new error probability model that includes false positive errors, resulting from decoding a non-transmitted codeword. The results show that the correlation allows for high reliability at the expense of spectral efficiency. This reflects the intuitive trade-off: an access from a massive number can be ultra-reliable only if the information across the devices is correlated. 
\end{abstract}

\IEEEpeerreviewmaketitle

\section{Introduction}
The interconnection of billions of devices within the Internet of Things (IoT) paradigm is one of the main challenges for future networks. Accordingly, the service structure of 5G, fully aligned with the ITU-R vision for IMT-2020, includes the massive Machine Type-Communication (mMTC) as one of the three core connectivity types. mMTC is typically defined through a scenario in which a massive number of IoT devices are connected to a Base Station (BS). The activation of the IoT devices is intermittent, such that at a given time, the IoT devices that are active and have a message to send constitute a random subset from the total set of devices~\cite{3GPPMachine2017}. A main use case for IoT is a distributed sensor network that intelligently monitors and manages a large number of devices~\cite{IEC2014}. The traffic in such systems can be (quasi-)periodic or event-driven~\cite{Nikaein2013}. In addition, source information and time correlations occur when many devices are sensing a common physical phenomenon. 

The conventional multiple access channel (MAC) has been well characterized~\cite{Plotnik1991, Ahlswede1971, Gallager1985}. The main results here are derived using the fact that the probability of successful joint decoding goes asymptotically to one with increasing blocklength. However, in the context of mMTC the devices have small data payloads. Even though a small subset of the devices are active simultaneously, the large total number of devices (up to \num{300000} in a single cell~\cite{Bockelmann2016}) means that the number of active devices can still be comparable to the blocklength. This results in finite blocklength (FBL) effects. A number of works have addressed the problem of massive access~\cite{Huang2012,Bockelmann2016}. However, in terms of theoretical rigor and fundamental results two works stand out, both of them assuming independent traffic. The first one is on the many-access channel by X. Chen et al.~\cite{Chen2017}. This paper shows the scaling of the number of users with the blocklength. On the other hand, Y. Polyanskiy provides a model in \cite{Polyanskiy2017} that is closer to the way massive access is commonly understood. Key elements of the model are devices employing the same codebook which precludes the identification of users and the error measure is done on a per-device basis. This has also been called unsourced random access \cite{Vem2017}. 

\begin{figure}[tb]
    \centering
    \includegraphics[width=0.60\textwidth]{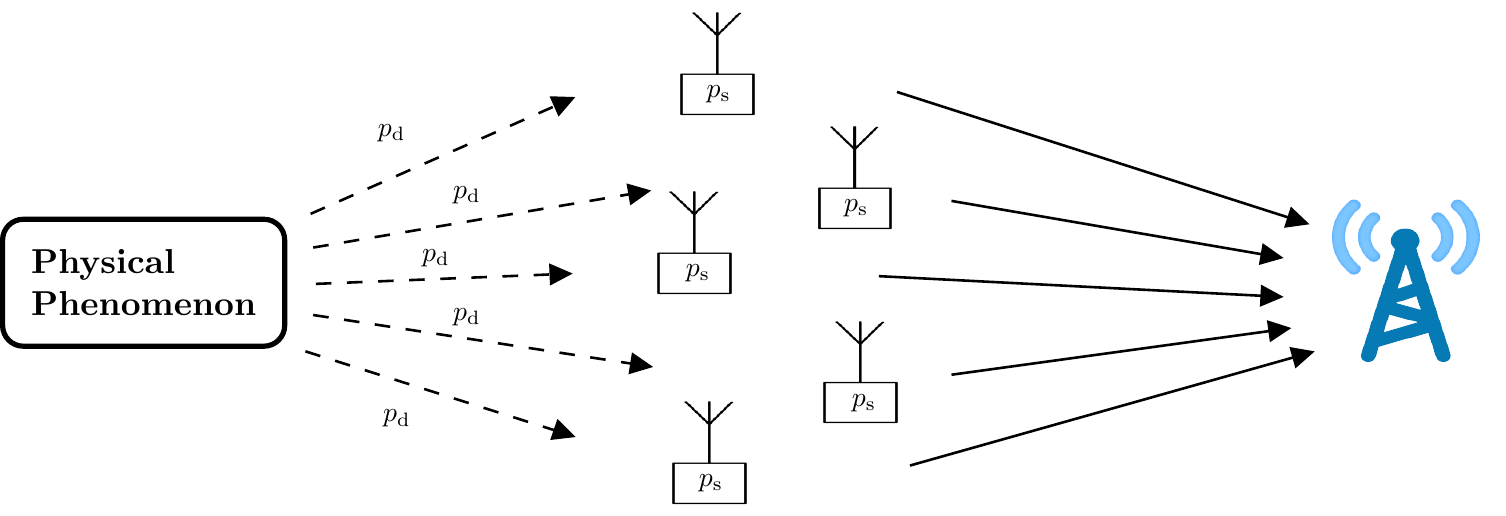}
    \caption{System model with common alarm and standard messages. $p_d$ denotes the probability of detecting an alarm, and $p_s$ is the probability of sending a standard message.}
    \label{fig:alarmSystem}
\end{figure}

In this we build upon the model in \cite{Polyanskiy2017} with an important extension: we bring in the correlation of activation and message content across different devices. This is different from the mainstream view on massive random access, where the device activation and message content is independent across the devices. 
An exemplary case is as follows: IoT devices can send standard messages or alarm messages, the latter with critical reliability requirement and triggered by a commonly observed phenomenon. In normal operation, standard uncorrelated messages are sent. Upon the alarm activation, a number of IoT devices will prioritize it and send the \emph{same} message. This reflects the extreme all-or-nothing correlation where devices are either mutually independent, or they are completely correlated both in source information and in time. Our model intends to capture the following intuitive observation. If the number of devices that transmit the same alarm message increases, then the reliability of the alarm message increases at the expense of the decrease of the total amount of information that comes from the total population of connected IoT devices. The model can be seen as having an (alarm) event that needs to be communicated through a random subset of devices, see Fig.~\ref{fig:alarmSystem}. By removing the alarm event the model boils down to the model in \cite{Polyanskiy2017}. 

Differently from previous works, the per-device probability of error is not meaningful for devices transmitting the alarm event in our model. Instead, the common alarm itself can be seen as a ``ghost'' device, which communicates through the actual IoT devices (see Fig.~\ref{fig:alarmSystem}) and we calculate the error probability with respect to this ghost device. In addition, the fact that we consider two message types (standard and alarm messages) necessitates the introduction of false positive errors, namely decoding a codeword that was not transmitted. In the system model in Fig.~\ref{fig:alarmSystem}, decoding an alarm message when no alarm has occurred is critical. This type of error is, typically, not  considered in a common communication-theoretic setting, where an error is defined as the event in which a decoder is not decoding a codeword correctly. 

The rest of the paper is organized as follows. Section \ref{sec:correlationModel} introduces the  system model including the source information and time correlations. In Section \ref{sec:spectralEfficiency} the entropy and the spectral efficiency of the correlated devices is derived. Section \ref{sec:ARA} defines the alarm random access code based on the novel error model, and the error bound is derived in Section \ref{sec:ErrorBound}. Finally, numerical evaluations are presented in Section \ref{sec:NumericalEvaluation}, and concluding remarks are given in Section \ref{sec:Conclusions}. Table \ref{tab:notation} lists the notation used in this paper.

 \section{Correlation Model}\label{sec:correlationModel}
\begin{table}[tb]
\centering
\resizebox{\columnwidth}{!}{
\centering
\begin{tabular}{l l| l l}
\toprule
\multicolumn{4}{c}{Notation}\\[.0\normalbaselineskip]
\midrule
$A$: & Alarm event &$\epsilon_{\mathrm{a}}$: & Target probability of error for alarm messages\\
$N$: & Total number of devices &$\epsilon_{\mathrm{sa}}$: & Target probability of error for standard messages in alarm event\\
$K$: & Number of active devices &$\epsilon_{\mathrm{fp}}$: & Target probability of false positive\\
$K_a$: & Number of devices sending an alarm message &$S$: & Spectral efficiency\\
$p_{\mathrm{a}}$: & Alarm probability &$H$: & Entropy\\
$p_{\mathrm{s}}$: & Standard message probability &$W_j$ & Message transmitted by the $j$-th device\\
$p_{\mathrm{d}}$: & Alarm detection probability &$a_i^j$ (lower case): & $(a_i, \ldots, a_j)$, $i\leq j$ for scalars/vectors\\
$\M_{\mathrm{s}}$: & Set of standard messages &$X_i^j$ (upper case): & $(X_i, \ldots, X_j)$, $i\leq j$ for random variables/vectors\\
$\M_{\mathrm{a}}$: & Set of alarm messages &$a_i^{i-1}$: & Empty tuple\\
$n$: & Blocklength &$\sum_{i=j}^{j-1}a_j$: & 0\\
$P'$: & Average transmission power &$[\mathcal{S}]^k$ & Set of all $k$-subsets of the set $\mathcal{S}$\\
$P$: & Maximal transmission power &$\mathcal{X}$: &$\in\mathbb{R}^n$ Input alphabet\\
$\epsilon_{\mathrm{s}}$: & Target probability of error for standard messages without an alarm & $\mathcal{Y}$: &$\in\mathbb{R}^n$ Output alphabet\\

\bottomrule
~
\end{tabular}
}
\caption{Notation used throughout this paper.}
\label{tab:notation}
\end{table}

We consider the uplink in a random access channel in which each access opportunity is a block of $n$ channel uses. In each block, $K$ out of $N$ devices transmit a message from one of the two disjoint message sets $\M_{\mathrm{s}}$ and $\M_{\mathrm{a}}$, consisting of $M_{\mathrm{s}}=|\M_{\mathrm{s}}|$ standard messages and $M_{\mathrm{a}}=|\M_{\mathrm{a}}|$ alarm messages, respectively. A typical case is having a stringent reliability requirement for the alarm messages, and a high throughput and massive access requirement for the rest. As also done in \cite{Polyanskiy2017}, we assume that the number of active devices, $K$, is known by the receiver.

Let $P_{\vec Y\vert \vec X_1^{K}}: [\mathcal{X}^n]^K\to\mathcal{Y}^n$ be a memoryless multiple access channel (MAC) satisfying permutation invariance where $\mathcal{X},\mathcal{Y}$ are the input and output alphabets. That is, the distribution $P_{\vec Y\vert \vec X_1^K}(\cdot\vert\vec x_1^K)$ coincides with $P_{\vec Y\vert \vec X_1^K}(\cdot\vert\vec x_{\pi(1)}, \ldots, \vec x_{\pi(K)})$ for any $\vec x_1^{K}\in[\mathcal{X}^n]^K$ and any permutation $\pi$. This assumption relates to the fact that no user identification is done at the receiver, i.e. unsourced random access~\cite{Vem2017}. Therefore, all devices use the same encoder $f:\mathcal{M}_{\mathrm{s}}\cup\mathcal{M}_{\mathrm{a}}\to \mathcal{X}^n$ and the receiver decodes according to the possibly randomized map $g:~\mathcal{Y}^n\to [\mathcal{M}_{\mathrm{s}}\cup\mathcal{M}_{\mathrm{a}}]^{K-K_{\mathrm{a}}+1}$, where $K_{\mathrm{a}}$ is the random number of devices that send alarm messages and $[\mathcal{S}]^k$ denotes the set of all $k$-subsets of the set $\mathcal{S}$. 

We denote the message transmitted by the $j$-th device as $W_j$. The transmitted messages are chosen according to the following model: An alarm event, $A$, occurs with probability $p_{\mathrm{a}}$, and there is no alarm with probability $1-p_{\mathrm{a}}$. If no alarm occurs then the system acts as in \cite{Polyanskiy2017}, i.e. each device transmits a message uniformly chosen from $\mathcal{M}_{\mathrm{s}}$ with probability $p_{\mathrm{s}}$, and it is silent with probability $1-p_{\mathrm{s}}$. If an alarm occurs, with probability $p_{\mathrm{d}}$ a device will detect it and transmit an alarm message. Contrary to the standard messages, all devices detecting the alarm send the \emph{same} message chosen uniformly from $\mathcal{M}_{\mathrm{a}}$. With probability $1-p_{\mathrm{d}}$ the device will act as if no alarm has occurred. It follows that $\mathbb{P}[W_j \in \mathcal{M}_{\mathrm{a}}] = p_{\mathrm{a}}p_{\mathrm{d}}$ and $\mathbb{P}[W_j \in \mathcal{M}_{\mathrm{s}}] = p_{\mathrm{s}} - p_{\mathrm{a}}p_{\mathrm{s}}p_{\mathrm{d}}$. Notice that the probability $p_{\mathrm{d}}$ in our model is the joint event of detecting an alarm and deciding to transmit a corresponding alarm message. The latter can be seen as a system design parameter and its impact to the system performance, particularly in the tradeoff between reliability and spectral efficiency, is discussed in next section. 

In contrast to practical random access scenarios, we assume that the number of active devices, $K$, is known by the receiver. This assumption can be justified by noting that $K$ could be estimated using the same procedure as in~\cite{Polyanskiy2017full}. Specifically, the base station can decode the received packets under the assumption that $K=0,1,\ldots,N$, and then re-generate the resulting packets and subtract them from the received signal. $K$ can then be determined based on the residual, which will equal the noise $\mathbf{Z}$, see (\ref{eq:NoiseDef}) if the correct value of $K$ has been determined. Furthermore, since the number of alarm messages, $K_{\mathrm{a}}$, is assumed unknown in the model, an incorrectly estimated $K$ will mainly affect the decoding of the non-critical standard messages. \section{Spectral efficiency}\label{sec:spectralEfficiency}
In this section, we study how the presence of common alarm messages affects the information transmitted in the system. We consider the system spectral efficiency defined as $S = \frac{H(W_1^K)}{n}$, where $K$ is the number of devices transmitting messages $W_1,...W_{K}$, $H$ is the joint entropy function and $n$ is the blocklength. 

The total number of devices, $N$, in the network affects the system spectral efficiency. To see this, consider the case with a high alarm detection probability $p_{\mathrm{d}}$, a low $p_{\mathrm{s}}$, alarm probability $p_{\mathrm{a}}=0.5$, and suppose we receive $10$ messages, i.e. $K=10$. If also $N=10$, then there is a high probability that an alarm has occurred since we know that all devices transmitted and that $p_{\mathrm{d}}$ is high. Moreover, in this case all devices have most likely transmitted the same message, resulting in a low spectral efficiency. On the other hand, with $N=10000$ devices in the network the probability that an alarm has occurred is low, being unlikely that $9990$ devices do not detect an alarm when $p_{\mathrm{d}}$ is high. In this case, the messages are likely to be distinct, resulting in a high spectral efficiency. 

The exact expression for the system spectral efficiency for this model is stated in Theorem~\ref{theo:jointEntropy}.
\begin{theo}\label{theo:jointEntropy}
For $K$ out of $N$ received messages and correlated devices as describe in Section~\ref{sec:correlationModel} the system spectral efficiency, $S$, is
\begin{equation}
S = \frac{1}{n}\sum_{k=1}^{K}H(W_k\vert W_{1}^{k-1}),
\end{equation}
where $H(W_k\vert W_1^{k-1})$ is given by 
\begin{align}
    \begin{split}
        H(W_{k}\vert W_1^{k-1}) =& ~ (B_0+B_1)~\sum_{i = 1}^{k-1}{k-1 \choose i}p_{\mathrm{a}}p_{\mathrm{d}}^i((1-p_{\mathrm{d}})p_{\mathrm{s}})^{k-1-i}N_0\\
                                                &-B_2\left(B_3\log_2\frac{B_3}{M_{\mathrm{a}}}+(1-B_3)\log_2\frac{1-B_3}{M_{\mathrm{s}}}\right),
   \end{split}
   \label{eq:expressionConditionalEntropy}
\end{align}
and
\begin{align}
    N_0 &= \frac{(p_{\mathrm{d}}+(1-p_{\mathrm{d}})p_{\mathrm{s}})^{K-(k-1)}(1-p_{\mathrm{d}})^{N-K}}{p_{\mathrm{a}}(p_{\mathrm{d}}+(1-p_{\mathrm{d}})p_{\mathrm{s}})^{K}(1-p_{\mathrm{d}})^{N-K} + (1-p_{\mathrm{a}})p_{\mathrm{s}}^{K}},\label{eq:N0}\\
    B_0 &= -\frac{p_{\mathrm{d}}}{p_{\mathrm{d}}+(1-p_{\mathrm{d}})p_{\mathrm{s}}}\log_2\left(\frac{p_{\mathrm{d}}}{p_{\mathrm{d}}+(1-p_{\mathrm{d}})p_{\mathrm{s}}}\right),\label{eq:B0}\\
    B_1 &= \frac{(1-p_{\mathrm{d}})p_{\mathrm{s}}}{p_{\mathrm{d}}+(1-p_{\mathrm{d}})p_{\mathrm{s}}}\left(\log_2 M_{\mathrm{s}}-\log_2\left(\frac{(1-p_{\mathrm{d}})p_{\mathrm{s}}}{p_{\mathrm{d}}+(1-p_{\mathrm{d}})p_{\mathrm{s}}}\right)\right),\label{eq:B1}\\
    B_2 &= \frac{p_{\mathrm{a}}(1-p_{\mathrm{d}})^{N-K+(k-1)}p_{\mathrm{s}}^{k-1}(p_{\mathrm{d}}+(1-p_{\mathrm{d}})p_{\mathrm{s}})^{K-(k-1)}+(1-p_{\mathrm{a}})p_{\mathrm{s}}^{K}}{p_{\mathrm{a}}(p_{\mathrm{d}}+(1-p_{\mathrm{d}})p_{\mathrm{s}})^{K}(1-p_{\mathrm{d}})^{N-K}+(1-p_{\mathrm{a}})p_{\mathrm{s}}^{K}},\label{eq:B2}\\
    B_3 &=\frac{p_{\mathrm{a}}p_{\mathrm{d}}(p_{\mathrm{d}}+(1-p_{\mathrm{d}})p_{\mathrm{s}})^{K-k}(1-p_{\mathrm{d}})^{N-K+k-1}p_{\mathrm{s}}^{k-1}}{p_{\mathrm{a}}(p_{\mathrm{d}}+(1-p_{\mathrm{d}})p_{\mathrm{s}})^{K-k+1}(1-p_{\mathrm{d}})^{N-K+k-1}p_{\mathrm{s}}^{k-1}+(1-p_{\mathrm{a}})p_{\mathrm{s}}^K}.\label{eq:B3}
\end{align}
\end{theo}
Proof of Theorem~\ref{theo:jointEntropy} can be found in Appendix~\ref{app:proofJointEntropy}.

For $p_{\mathrm{a}}=0$ or $p_{\mathrm{d}}=0$ (i.e. no correlation) the system spectral efficiency is the well-known $\frac{K}{n}\log_2M_\mathrm{s}$ as in~\cite{Polyanskiy2017}.

 \section{Alarm Random Access Codes} \label{sec:ARA}
We now define a random access code that allows for reliability diversity for standard and alarm messages. This entails having different error events for the two message types. Specifically, in order to capture the characteristics of alarm messages, we introduce reliability constraints that relates to the certainty of decoding alarm messages in the event of an alarm, but also to the certainty of \emph{not} decoding alarm messages when no alarms has occurred (false positives). This is different from usual analysis since we need not only consider the event of incorrectly decoding a message, but also the type of message that is decoded instead. 

The error events are listed in Table \ref{tab:errorEvents}, where we have included the ``No error" column to emphasize the opposite characteristics of alarm messages and standard messages.
We define error events for standard messages as in \cite{Polyanskiy2017}, i.e. errors are considered per-device and the event that more than one device sends the same standard message results in an error. In contrast, no error occurs if multiple devices transmit the same alarm message. Similarly, decoding distinct alarm messages also results in an error since only one alarm is assumed to be active at a time, while decoding distinct standard messages is naturally not an error. Formally, we define the following error events: $E_j\triangleq\{W_j\notin g(\vec Y)\}\cup\{W_j=W_i \text{ for some } i\neq j\}$ is the event of not decoding the message from the $j$-th device, $E_{\mathrm{a}}\triangleq \{W_0\notin g(\vec Y)\}\cup\{\vert g(\vec Y)\cap \M_{\mathrm{a}}\vert>1\}$ for $W_0\in\M_{\mathrm{a}}$ is the event of not decoding an alarm message or decoding more than one, and $E_{\mathrm{fp}}\triangleq\{g(\vec Y)\cap\M_{\mathrm{a}}\neq \varnothing\}$ is the event of decoding any alarm message (which is an error when no alarm has occurred). This leads to the following definition of a $K$-user alarm random access (ARA) code.
\begin{table}[tb]
\centering
\resizebox{0.7\columnwidth}{!}{
\centering
\begin{tabular}{l l l}
\toprule
\multicolumn{3}{c}{Classification of events}\\[.0\normalbaselineskip]
& \qquad\qquad \textbf{Error} & \qquad\qquad \textbf{No error}\\
\midrule
\textbf{No alarm} & \llap{-} A standard message is not decoded: &  \llap{-} A standard message is decoded:\\ 
                  & \qquad$\{\mathcal{M}_{\mathrm{s}}\ni W_j\notin g(\vec Y)\}$ & \qquad$\{\mathcal{M}_{\mathrm{s}}\ni W_j\in g(\vec Y)\}$\\[.0\normalbaselineskip]
         & \llap{-} More than one device sends the same message: & \llap{-} Different messages are sent:  \\
         & \qquad$\{W_j = W_i \text{ for some } i\neq j\}$ & \qquad $\{W_j\neq W_i ~\forall~ i\neq j\}$\\[.0\normalbaselineskip]
         & \llap{-} At least one alarm message is decoded:& \llap{-} No alarm message is decoded:\\
         & \qquad$\{g(\vec Y)\cap \mathcal{M}_{\mathrm{a}} \neq \varnothing\}$ \quad (false positive) & \qquad$\{g(\vec Y)\cap\mathcal{M}_{\mathrm{a}} = \varnothing\}$\quad (true negative) \\[.0\normalbaselineskip]
\midrule
\textbf{Alarm} & \llap{-} The alarm message is not decoded: & \llap{-} The alarm message is decoded:\\
               &\qquad $\{W_0\notin g(\vec Y)\}$\quad (false  negative) & \qquad $\{W_0\in g(\vec Y)\}$\quad (true positive)\\[.0\normalbaselineskip]
      & \llap{-} More than one alarm message is decoded: & \llap{-} More than one device sends the same alarm message:\\
      & \qquad $\{\vert g(\vec Y)\cap\mathcal{M}_{\mathrm{a}}\vert>1\}$  & \qquad $\{W_j = W_i=W_0\in\mathcal{M}_{\mathrm{a}} \text{ for some } i\neq j\}$\\[.0\normalbaselineskip]
      & \llap{-} A standard message is not decoded: &\\
      & \qquad $\{\mathcal{M}_{\mathrm{s}}\ni W_j\notin g(\vec Y)\}$ &\\[0.0\normalbaselineskip]
      & \llap{-} Two or more device sends the same standard &\\
      & ~message: & \\
      & \qquad $\{W_i = W_j \in\mathcal{M}_{\mathrm{s}}\text{ for some } i\neq j\}$ & \\
\bottomrule
~
\end{tabular}
}
\caption{Error events in the considered system. In the alarm event we denote the alarm message by $W_0$.}
\label{tab:errorEvents}
\end{table}

\begin{defi}\label{def:ARAcode}
    An $(M_{\mathrm{s}}, M_{\mathrm{a}}, n, \epsilon_{\mathrm{a}}, \epsilon_{\mathrm{s}},\epsilon_{\mathrm{sa}}, \epsilon_{\mathrm{fp}})$ alarm random access (ARA) code for the $K$-user channel $P_{\vec Y\vert \vec X_1^{K}}$ is a pair of (possibly randomized) maps, the encoder $f:\M_{\mathrm{s}}\cup\M_{\mathrm{a}}\to\mathcal{X}^n$, and the decoder $g:\mathcal{Y}^n\to[\M_{\mathrm{s}}\cup\M_{\mathrm{a}}]^{K-K_{\mathrm{a}}+1}$ satisfying 
    \begin{align}
        \P{E_{\mathrm{a}}\vert A} &\leq \epsilon_{\mathrm{a}},\label{eq:def_eps_a}\\
        \frac{1}{K}\sum_{j=1}^{K}\P{E_j\vert\neg A}&\leq \epsilon_{\mathrm{s}},\label{eq:def_eps_s}\\
        \E[K_{\mathrm{a}}]{\frac{1}{K-K_{\mathrm{a}}}\sum_{j=1}^{K-K_{\mathrm{a}}}\P{E_j\vert A}}&\leq\epsilon_{\mathrm{sa}},\label{eq:def_eps_sa}\\
        \P{E_{\mathrm{fp}}\vert \neg A}&\leq\epsilon_{\mathrm{fp}},\label{eq:def_eps_fp}
    \end{align}
    where $\vec X_j = f(W_j)$,  $W_1, \ldots, W_K \in\M_{\mathrm{s}}$ when there is no alarm and $W_1,\ldots,W_{K-K_{\mathrm{a}}}\in\M_{\mathrm{s}}$, $W_{K-K_{\mathrm{a}}+1}=\ldots=W_K=W_0\in\M_{\mathrm{a}}$ in the alarm event for a random number, $K_{\mathrm{a}}$, alarm messages. 
\end{defi}
The left hand side of \eqref{eq:def_eps_a} is the probability of not decoding or resolving the alarm message in the alarm event. The left hand side of \eqref{eq:def_eps_s} is the average per-device error probability when there is no alarm, and \eqref{eq:def_eps_sa} refers to the case when there is an alarm. Lastly left hand side of \eqref{eq:def_eps_fp} is the probability of false positives. In a practical scenario the entities $\epsilon_{\mathrm{a}}$, $\epsilon_{\mathrm{s}}$, $\epsilon_{\mathrm{sa}}$ and $\epsilon_{\mathrm{fp}}$ can be treated as reliability requirements, in which case the achievability of an ARA code is of interest.

In the remainder of the paper we limit the analysis to the Gaussian MAC (GMAC) given by 
\begin{equation}\label{eq:NoiseDef}
    \vec Y= \sum_{m=1}^{K}\vec X_m + \vec Z.
\end{equation}
where $\vec Z\in\mathbb{R}^{n}$ is a standard Gaussian noise vector. Additionally a maximal average transmission power, $P$, is included. That is we require $\norm{f(W_j)}_2\leq nP$.
This model is based on the assumption that the blocklength is short enough to be within the coherence time of the channel. This allows for the devices to do channel inversion and precode their signals so that they add up coherently at the receiver. This gives the possibility of a very high reliability for alarm messages.
 \section{Random Coding Error Bound} \label{sec:ErrorBound}
The achievability conditions for an ARA code are presented in Theorem~\ref{theo:RandomCodingBound}, which provides bounds for the error probabilities $\epsilon_{\mathrm{a}}$, $\epsilon_{\mathrm{s}}$, $\epsilon_{\mathrm{sa}}$ and $\epsilon_{\mathrm{fp}}$ for a given blocklenght $n$, message set sizes $M_{\mathrm{a}}$ and $M_{\mathrm{s}}$, average transmission power $P'$, and maximal transmission power $P$. 
\begin{theo}\label{theo:RandomCodingBound}
    Fix $P'<P$. There exists an $(M_{\mathrm{a}}, M_{\mathrm{s}}, n, \epsilon_{\mathrm{a}}, \epsilon_{\mathrm{s}}, \epsilon_{\mathrm{sa}}, \epsilon_{\mathrm{fp}})$ alarm random access code for the $K$-user GMAC satisfying power-constraint $P$ and 
    \begin{align}
        \epsilon_\mathrm{a}&\leq \sum_{K_{\mathrm{a}}=0}^{K}p_{K_{\mathrm{a}}}(K_{\mathrm{a}})a(K,K_{\mathrm{a}}) + p_0,\label{eq:alarmErrorBound}\\
        \epsilon_\mathrm{s}&\leq b(K)+ c(K) - b(K)c(K),\label{eq:noAlarmErrorBound}\\
        \epsilon_{\mathrm{sa}} &\leq \sum_{K_{\mathrm{a}}=0}^{K}p_{K_{\mathrm{a}}}(K_{\mathrm{a}})\left(1-d\left(K,K_{\mathrm{a}}\right)\left(1-c\left(K-K_{\mathrm{a}}\right)\right)\right)\label{eq:AlarmStdErrorBound}\\
        \epsilon_{\mathrm{fp}} &\leq b(K).\label{eq:fPErrorBound}
    \end{align}
Defining $\phi(k, \alpha) = \frac{1}{2}\ln(1+2kP'\alpha)$ and $\Phi(k,\alpha) = \frac{\alpha}{1+2kP'\alpha}$, then related to \eqref{eq:alarmErrorBound}:
\begin{align}
    p_{K_{\mathrm{a}}}(k) &= {K \choose k} \frac{p_{\mathrm{d}}^k\left(\left(1-p_{\mathrm{d}}\right)p_{\mathrm{s}}\right)^{K-k}}{(p_{\mathrm{d}} + (1-p_{\mathrm{d}})p_{\mathrm{s}})^{K}}, \label{eq:distKcondA}\\
    a(K,K_{\mathrm{a}})&= \min\left(\sum_{K_{\mathrm{a}}'=0}^{K}e^{-nE_{\mathrm{a}}}, ~1\right),\label{eq:a}\\
    p_0 &=  \P{\frac{1}{n}\sum_{i=1}^{n}Z_i^2> \frac{P}{P'}},\\
    E_{\mathrm{a}}&=\max_{0\leq\rho\leq 1, 0<\lambda_{\mathrm{a}}} -\frac{\rho}{n}\ln(M_\mathrm{a}-1)+\xi_{\mathrm{a}},\\
    \xi_{\mathrm{a}}&=\rho \phi(K_{\mathrm{a}}'^2, \lambda_{\mathrm{a}})+\phi(K_{\mathrm{a}}^2, \rho\beta_{\mathrm{a}}) + \phi(K-K_{\mathrm{a}}, \gamma_{\mathrm{a}})+\phi(1/P',\psi_{\mathrm{a}}),\label{eq:xiA}\\
    \psi_{\mathrm{a}} &= \Phi(K-K_{\mathrm{a}}, \gamma_{\mathrm{a}}), ~~\gamma_{\mathrm{a}}=\Phi(K_{\mathrm{a}}^2,\rho\beta_{\mathrm{a}})-\rho\lambda_{\mathrm{a}}, ~~\beta_{\mathrm{a}}=\Phi(K_{\mathrm{a}}'^2,\lambda_{\mathrm{a}}).\label{eq:psiGammaAndBeta}
\end{align}
Related to \eqref{eq:fPErrorBound}:
\begin{align}
    b(K) &= \min\left(\sum_{K_{\mathrm{a}}'=1}^{K}e^{-nE_{\mathrm{fp}}},~1\right),\label{eq:b}\\
    E_{\mathrm{fp}} &= \max_{0\leq\rho\leq 1,~0<\lambda_{\mathrm{fp}}} -\frac{\rho}{n}\ln(M_{\mathrm{a}})+\xi_{\mathrm{fp}},\\
    \xi_{\mathrm{fp}} &= \rho \phi(K_{\mathrm{a}}'^2, \lambda_{\mathrm{fp}}) + \phi(K, \rho\beta_{\mathrm{fp}}) + \phi(1/P', \gamma_{\mathrm{fp}}),\\
    \gamma_{\mathrm{fp}} &= \Phi(K,\rho\beta_{\mathrm{fp}}),~~ \beta_{\mathrm{fp}} = \Phi(K_{\mathrm{a}}'^2, \lambda_{\mathrm{fp}})-\lambda_{\mathrm{fp}},\label{eq:gammaBetaFp}
\end{align}
Related to \eqref{eq:noAlarmErrorBound}
\begin{align}
    c(K) &= \sum_{t=1}^{K}\frac{t}{K}\min(p_t, q_t)+\frac{{K\choose 2}}{M_{\mathrm{s}}}+Kp_0,\label{eq:yuriErrorBound}\\
    p_t &= e^{-nE_t},\\
    E_t &= \max_{0\leq\rho,\rho_1\leq 1}-\rho\rho_1tR_1-\rho_1R_2 + E_0(\rho,\rho_1),\\
    E_0(\rho,\rho_1) &= \rho\rho_1\phi(t,\lambda_{\mathrm{s}}) + \rho_1\phi(t,\mu)+\frac{1}{2}\ln(1-2b\rho_1),\\
    b &= \rho\lambda_{\mathrm{s}}-\Phi(t, \mu), ~~\mu = \rho\Phi(t, \lambda_{\mathrm{s}}),~~ \lambda_{\mathrm{s}} = \frac{P't-1+\sqrt{D}}{4(1+\rho_1\rho)P't},\\
    D &= (P't-1)^2 + 4P't\frac{1+\rho\rho_1}{1+\rho},\\
    R_1 &= \frac{1}{n}\ln(M_s)-\frac{1}{nt}\ln(t!),~~ R_2 = \frac{1}{n}\ln{K\choose t},\\
    q_t &= \inf_{\gamma_{\mathrm{s}}} \P{I_t\leq\gamma_{\mathrm{s}}}+e^{n(tR_1+R_2)-\gamma_{\mathrm{s}}},\\
    I_t &= \min_{S_0\in[\M_{\mathrm{s}}]^t}i_t\left(\sum_{W\in S_0}\vec c_W;\vec Y\vert \sum_{W\in S_0^c}\vec c_W)\right),\\
    i_t(\vec a;\vec y\vert \vec b) &= nC_t+\frac{\ln e}{2}\left(\frac{\norm{\vec y-\vec b}_2}{1+P't}-\norm{\vec y-\vec a-\vec b}_2\right),
\end{align}
where $C_t = \phi(1/2, t)$, $S_0\in[\M_{\mathrm{s}}]^t$ is a $t$-subset of true standard messages and $\vec c_{W}\sim\mathcal{N}(\vec 0, \vec I_nP')$ is codeword corresponding to message $W$.
Related to \eqref{eq:AlarmStdErrorBound}:
\begin{align}
    d(K, K_{\mathrm{a}}) &= (1-(a(K,K_{\mathrm{a}})+p_0))(1-e(K,K_{\mathrm{a}})+p_0),\\
    e(K,K_{\mathrm{a}}) &= \min\left(\sum_{\substack{K_{\mathrm{a}}'=0\\K_{\mathrm{a}}'\neq K_{\mathrm{a}}}}^{K}e^{-nE_{\mathrm{sa}}},~1\right),\label{eq:e}\\
    E_{\mathrm{sa}} &= \max_{0<\lambda_{\mathrm{sa}}} \phi((K_{\mathrm{a}}-K_{\mathrm{a}}')^2, \lambda_{\mathrm{sa}}) + \phi(K-K_{\mathrm{a}}, \beta_{\mathrm{sa}}) + \phi(1/P', \gamma_{\mathrm{sa}}),\label{eq:xiSa}\\
    \gamma_{\mathrm{sa}} &= \Phi\left(K-K_{\mathrm{a}}, \beta_{\mathrm{sa}}\right),\quad \beta_{\mathrm{sa}} = \Phi\left((K_{\mathrm{a}}-K_{\mathrm{a}}')^2, \lambda_{\mathrm{sa}}\right)-\lambda_{\mathrm{sa}}.\label{eq:gammaBetaSa}
\end{align}
\end{theo}
Proof of Theorem~\ref{theo:RandomCodingBound} can be found in Appendix~\ref{app:proofRandomCodingBound}.

 \section{Numerical evaluation} \label{sec:NumericalEvaluation}
The bounds in Theorem~\ref{theo:RandomCodingBound} are given for a fixed number of active devices, $K$, but the probability of a given value of $K$ depends on whether an alarm has happened or not. Therefore, we consider the average bound over the distribution of $K$ conditioned on the alarm state and the total number of devices, $N$. The distribution of $K$ given an alarm is 
\begin{equation}
    p_K(k\vert A) = {N \choose k}(p_{\mathrm{d}}+(1-p_{\mathrm{d}})p_{\mathrm{s}})^k(1-p_{\mathrm{d}})^{N-k}(1-p_{\mathrm{s}})^{N-k},
\end{equation}
and the distribution of $K$ given no alarm is 
\begin{equation}
    p_K(k\vert\neg A) = {N \choose k}p_{\mathrm{s}}^{k}(1-p_{\mathrm{s}})^{N-k}.
\end{equation}

We first study the trade-off between the probability of error for alarm messages and the per-device spectral efficiency $S$, during the event of an alarm. We consider a setting with $N=\num{1000}$ devices and a blocklength of $n=\num{30000}$. The alarm and standard messages are $3$ and $\num{100}$ bits, respectively. The probability of activation when there is no alarm is $p_{\mathrm{s}}=0.01$, and the transmission power is chosen such that the target average error bound for standard messages is $\epsilon_{\mathrm{s}} = \num{e-1}$, and the probability of false positive alarms is below $\epsilon_{\mathrm{fp}} = \num{e-5}$. Having only a few bits for alarm messages is a realistic setting, e.g. in a sensor network the alarm could be that a sensed value is too high or too low resulting in only one bit needed for the alarm message.
\begin{figure}[tb]
    \centering
    \includegraphics[width=0.60\textwidth]{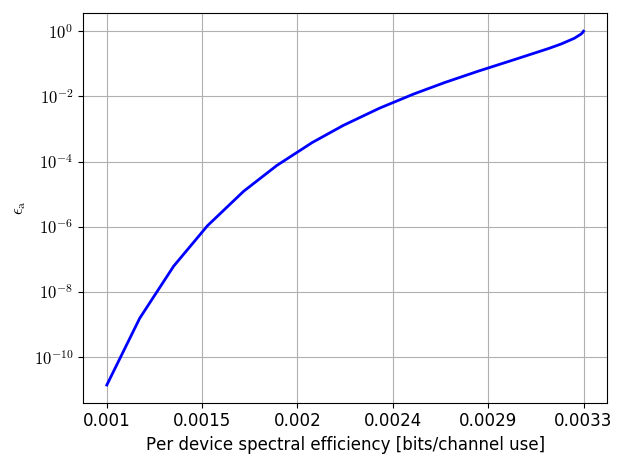}
    \caption{Trade-off between probability of error for alarm messages and the spectral efficiency.  Blocklength $n=\num{30000}$, $N=\num{1000}$, target error probabilities $\epsilon_{\mathrm{s}} = \num{e-1}$, $\epsilon_{\mathrm{fp}} = \num{e-5}$, set sizes $M_{\mathrm{s}}=2^{100}$, $M_{\mathrm{a}}=2^3$, $p_{\mathrm{s}}=0.01$ and $p_{\mathrm{a}}=1$.}
    \label{fig:errorVsSpecEff}
\end{figure}

In Fig.~\ref{fig:errorVsSpecEff} it can be seen that the probability of error increases for increasing spectral efficiency (decreasing $p_{\mathrm{d}}$). Notice that the maximum spectral efficiency is achieved when the error probability is one (or equivalently, $p_{\mathrm{d}}=0$), i.e. no alarm messages are detected. This is expected since a higher number of devices transmitting alarm messages reduces the per-device spectral efficiency, but increases the received signal-to-noise ratio of alarm messages. Furthermore, very high reliability is achievable. This trade-off between spectral efficiency and probability of error is not surprising since this is also the case when the blocklength or message set size are changed. The novelty is in the fact that it is the correlation between devices that causes the trade-off.

We now consider the minimal average transmission power, $P'$, required to satisfy some target error probabilities. We assume no power restriction. That is $p_0 = 0$ in Theorem~\ref{theo:RandomCodingBound}. Let all parameters be fixed except $P'$ and $p_{\mathrm{d}}$. This is an optimization problem on the form
\begin{equation}
   \begin{array}{lll}
       \displaystyle  \minimize_{0\leq P',~0\leq p_{\mathrm{d}}\leq 1} & P'& \\
       \mbox{s.t.}  & \sum_{K=0}^{N}p_K(K\vert \neg A)(b(K)+c(K)-b(K)c(K))&\leq \epsilon_{\mathrm{s}}\\
                    & \sum_{K=0}^{N}p_K(K\vert\neg A)b(K)&\leq \epsilon_{\mathrm{fp}}\\
                    & \sum_{K=0}^{N}p_K(K\vert A)\sum_{K_{\mathrm{a}}=0}^{K}p_{K_{\mathrm{a}}}(K_{\mathrm{a}})a(K,K_{\mathrm{a}})&\leq \epsilon_{\mathrm{a}}\\
                    & \sum_{K=0}^{N}p_K(K\vert A)\sum_{K_{\mathrm{a}}=0}^{K}p_{K_{\mathrm{a}}}(K_{\mathrm{a}})(1-d(K,K_{\mathrm{a}})(1-c(K-K_{\mathrm{a}})))&\leq \epsilon_{\mathrm{sa}}
   \end{array}\label{eq:optProblem}
\end{equation}
The constraint functions are strictly increasing for decreasing values of $P'$ and $p_{\mathrm{d}}$, and the first two constraints do not depend on $p_{\mathrm{d}}$. Therefore, the problem can be efficiently solved using bisection by first minimizing $P'$ subject to the first two constraints, and then determine $p_{\mathrm{d}}$ from the last two constraints using the $P'$ from the previous step. This provides a feasible solution for the values of $N, \epsilon_{\mathrm{a}}$ and $\epsilon_{\mathrm{sa}}$ of interest in this paper. However, if no feasible $p_{\mathrm{d}}\in[0,1]$ can be obtained from the second step, then a solution can be found by setting $p_{\mathrm{d}}=1$ and minimizing $P'$ subject to the last two constraints.

We use the same system parameters as in the previous scenario, except that we now fix $\epsilon_{\mathrm{a}} = \epsilon_{\mathrm{fp}} = \num{e-5}$ and $\epsilon_{\mathrm{s}}=\epsilon_{\mathrm{sa}} = \num{e-1}$. Based on the optimal $p_{\mathrm{d}}$ and the values of $p_{\mathrm{s}}$, $p_{\mathrm{a}}$, we evaluate the minimal average energy-per-bit $\E[p_K]{\frac{E_0}{N_0}}=\frac{nP'}{2\E[p_K]{H(W_1^K)/K}}$. 

\begin{figure}[tb]
    \centering
    \includegraphics[width=0.60\textwidth]{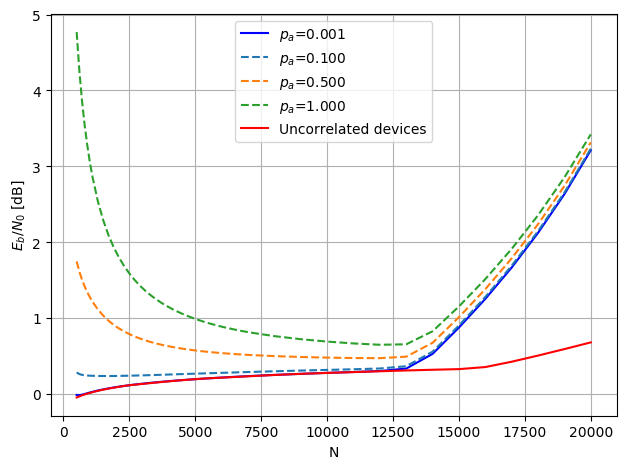}
    \caption{Trade-off between $\frac{E_b}{N_0}$ and the number of devices, $N$, for different values of alarm probability $p_{\mathrm{a}}$ and for uncorrelated devices. Blocklength $n=\num{30000}$, target error probabilities $\epsilon_{\mathrm{a}} = \epsilon_{\mathrm{fp}} = \num{e-5}$, $\epsilon_{\mathrm{s}}=\epsilon_{\mathrm{sa}} = \num{e-1}$, set sizes $M_{\mathrm{s}}=2^{100}$, $M_{\mathrm{a}}=2^3$ and $p_{\mathrm{s}}=0.01$.}
    \label{fig:EnergyPerBitVarPaN=500-20000}
\end{figure}

In Fig.~\ref{fig:EnergyPerBitVarPaN=500-20000} the solid blue line shows the energy-per-bit as a function of total number devices, $N$, for this setup. Notice that optimization is done for each $N$. Additionally, the achievable energy-per-bit for the uncorrelated case $(p_{\mathrm{d}}=0)$ is included for reference, and is obtained as described in \cite{Polyanskiy2017} but without the transmission power restriction.
It can be seen that almost the same energy-per-bit is achievable for correlated and uncorrelated devices up to approximately $13000$ devices, where the energy-per-bit required in the correlated case starts to increase significantly. This is due to the fact that the bound for false positives starts to dominate the choice of $P'$. Thus, due to high multi-access interference, the probability of decoding a false positive is higher than the probability of failing to decode a standard message. The increase in power needed to accommodate the false positive target probability $\epsilon_{\mathrm{fp}}$ causes the error probability of standard messages to go well below their target error probability $\epsilon_{\mathrm{s}}$. In fact, with more than $17500$ devices the error bound for standard messages is also approximately \num{e-5}. This is similar to the behavior in the uncorrelated case where the finite blocklength penalty is the dominating constraint when $N$ is small, while multi-access interference dominates for large $N$~\cite{Polyanskiy2017}. This is seen in the increase in the slope at around $15000$ devices in the uncorrelated case.

The effect of increasing alarm probability, $p_{\mathrm{a}}$, can be seen as the dashed curves in Fig.~\ref{fig:EnergyPerBitVarPaN=500-20000}. The energy-per-bit is higher for larger $p_{\mathrm{a}}$ due to the increased rate of alarm events where spectral efficiency is lower. The energy-per-bit in alarm events corresponds to the curve for $p_a=1$. Notice that the energy requirement $P'$ and the probability $p_{\mathrm{d}}$ are not altered by varying $p_{\mathrm{a}}$ since the error probabilities for ARA codes are conditioned on the occurrence of an alarm. The high energy-per-bit for small $N$ and high $p_{\mathrm{a}}$ is due to the large number of devices (relative to $N$) that must devote their resources to a single alarm message in order to accommodate the target alarm reliability. In general, the curves corresponding to different values of $p_{\mathrm{a}}$ are approaching each other for increasing $N$. This is caused by the almost constant number of alarm messages required to achieve the alarm target reliability. For increasing $N$ this ratio of alarm messages to standard messages grows and the traffic will be mostly standard messages.
 \section{Conclusions} \label{sec:Conclusions}
We have studied the trade-off between reliability and spectral efficiency in a massive random access scenario where the devices can send standard messages or alarm messages. The alarm messages are triggered by a common physical phenomenon and introduce correlation in both the transmitted messages and the activation of devices.
We derive the system spectral efficiency and propose an achievability bound for alarm random access codes. We show that very reliable transmissions of alarm messages can be achieved, but that the correlation causes a trade-off in spectral efficiency. In particular, when the multi-access interference is moderate, the cost of providing high reliability of alarm messages is small in terms of the average energy-per-bit. However, when multi-access interference is high, the probability of decoding a false positive alarm message dominates the error probabilities, and the cost of providing high reliability is significant.

\appendices
\section{Proof of Theorem~\ref{theo:jointEntropy}}\label{app:proofJointEntropy}
To explicitly show the dependency on the number of messages define 
\begin{equation}
    T_{K}^N = \{W_1\in\mathcal{M}_{\mathrm{a}}\cup\mathcal{M}_{\mathrm{s}}\}\cup\cdots\cup\{W_{K}\in\mathcal{M}_{\mathrm{a}}\cup\mathcal{M}_{\mathrm{s}}\}\cup\{W_{K+1}\in\varnothing\}\cup\cdots\cup\{W_N\in\varnothing\}
\end{equation}
as the event that the first $K$ out of $N$ devices transmit and the rest are silent. Due to symmetry in the devices, and without loss of generality, we assume the $K$ first devices that are transmitting. By the law of total probability this event has probability 
\begin{equation}
    p(T_{K}^{N}) = p_{\mathrm{a}}(p_{\mathrm{d}}+(1-p_{\mathrm{d}})p_{\mathrm{s}})^{K}(1-p_{\mathrm{d}})^{N-K}(1-p_{\mathrm{s}})^{N-K} + (1-p_{\mathrm{a}})p_{\mathrm{s}}^{K}(1-p_{\mathrm{s}})^{N-K},
   \label{eq:probT}
\end{equation}

System spectral efficiency, $S$, is defined as $S = H(W_1^K)/n$ where the joint entropy of all $K$ messages can be expressed using the chain rule for entropy~\cite[Theo. 2.5.1]{Cover2006} as
\begin{equation}
H(W_1^{K}) = \sum_{k=1}^{K}H(W_k\vert W_{1}^{k-1}).
    \label{eq:chainRuleEntropy}
\end{equation}
Thus we need to express the conditional entropy $H(W_k\vert W_1^{k-1})$ given by
\begin{equation}
    H(W_k\vert W_1^{k-1}) = \sum_{w_1\in\mathcal{M}_{\mathrm{a}}\cup\mathcal{M}_{\mathrm{s}}}\cdots\sum_{w_{k-1}\in\mathcal{M}_{\mathrm{a}}\cup\mathcal{M}_{\mathrm{s}}}
    \!p(w_1^{k-1}\vert T_{K}^{N})H(W_{k}\vert W_{1}^{k-1}=w_{1}^{k-1}),
    \label{eq:generalCondEntropy}
\end{equation}
where 
\begin{equation}
    H(W_{k}\vert W_1^{k-1} = w_{1}^{k-1}) = -\sum_{w_{k}\in\mathcal{M}_{\mathrm{a}}\cup\mathcal{M}_{\mathrm{s}}}p(w_{k}\vert w_1^{k-1},T_{K}^{N})\log_2(p(w_{k}\vert w_1^{k-1}, T_{K}^{N})),
    \label{eq:condEntropyGivenw}
\end{equation}
for $k\leq K$.

Observe that $\M_{\mathrm{s}}$ and $\M_{\mathrm{a}}$ are disjoint so that we can split each sum in \eqref{eq:generalCondEntropy} into two sums over $w_i\in\M_{\mathrm{a}}$ and $w_i\in\M_{\mathrm{s}}$.
For convenience, we define the set $A^{k} = \{w_1^k~\vert ~w_1^k\in[\M_{\mathrm{a}}\cup\M_{\mathrm{s}}]^k, \exists~0\leq i\leq k: w_i\in\M_{\mathrm{a}}\}$ as the set of $k$-subsets of $\M_{\mathrm{a}}\cup\M_{\mathrm{s}}$ that contain at least one alarm message and rewrite \eqref{eq:generalCondEntropy} as
\begin{equation}
    H(W_k\vert W_1^{k-1}) = \sum_{w_1^{k-1}\in A^{k-1}} p_{\mathrm{A}}(w_1^{k-1}\vert T_K^N)H_{\mathrm{A}}(W_k\vert W_1^{k-1}=w_1^{k-1}) + \sum_{w_1^{k-1}\in[\M_{\mathrm{s}}]^{k-1}}p_{\mathrm{S}}(w_1^{k-1}\vert T_K^N)H_{\mathrm{S}}(W_k\vert W_1^{k-1}=w_1^{k-1}).
    \label{eq:generalCondEntSplit}
\end{equation}

We first derive an expression for $H_{\mathrm{A}}(W_k\vert W_1^{k-1}=w_1^{k-1})$ using the fact that at least one of $w_1,\ldots,w_{k-1}$ belongs to $\M_{\mathrm{a}}$. We additionally split the sum in \eqref{eq:condEntropyGivenw} into two sums; one over $w_k\in\M_{\mathrm{a}}$ and one over $w_k\in\M_{\mathrm{s}}$:
 \begin{align}
     \begin{split}
        H_{\mathrm{A}}(W_{k}\vert W_1^{k-1} = w_{1}^{k-1}) &= -\sum_{w_{k}\in\mathcal{M}_{\mathrm{a}}}p(w_{k}\vert w_1^{k-1}\in A^{k-1},T_{K}^{N})\log_2(p(w_{k}\vert w_1^{k-1}\in A^{k-1}, T_{K}^{N}))\\
        &\quad-\sum_{w_{k}\in\mathcal{M}_{\mathrm{s}}}p(w_{k}\vert w_1^{k-1}\in A^{k-1},T_{K}^{N})\log_2(p(w_{k}\vert w_1^{k-1}\in A^{k-1}, T_{K}^{N})).
        \label{eq:condEntropyGivenw2}
    \end{split}
\end{align}
Using Bayes' theorem we obtain
\begin{align}
    p(W_k\in\mathcal{M}_{\mathrm{a}}\vert w_1^{k-1}\in A^{k-1}, T_{K}^{N}) &= \frac{p(T_{K}^{N}\vert W_k\in\mathcal{M}_{\mathrm{a}}, w_1^{k-1}\in A^{k-1})p(W_k\in\mathcal{M}_{\mathrm{a}}\vert w_1^{k-1}\in A^{k-1})}{p(T_{K}^{N}\vert w_1^{k-1}\in A^{k-1})}\nonumber\\
                                                                &= \frac{(p_{\mathrm{d}}+(1-p_{\mathrm{d}})p_{\mathrm{s}})^{K-k}(1-p_{\mathrm{d}})^{N-k}(1-p_{\mathrm{s}})^{N-k}p_{\mathrm{d}}}{(p_{\mathrm{d}}+(1-p_{\mathrm{d}})p_{\mathrm{s}})^{K-(k-1)}(1-p_{\mathrm{s}})^{N-k}(1-p_{\mathrm{s}})^{N-k}}\nonumber\\
                                                                &= \frac{p_{\mathrm{d}}}{p_{\mathrm{d}}+(1-p_{\mathrm{d}})p_{\mathrm{s}}}\label{eq:pWinMaGivenAT}.
\end{align}
Since all devices that detect the alarm transmit the same message, the first term in \eqref{eq:condEntropyGivenw2} is
\begin{align}
    -\sum_{w_k\in\M_{\mathrm{a}}}p(w_k\vert w_1^{k-1}\in A^{k-1},T_K^N)\log_2(p(w_k\vert w_1^{k-1}\in A^{k-1},T_K^N))&=-\frac{p_{\mathrm{d}}}{p_{\mathrm{d}}+(1-p_{\mathrm{d}})p_{\mathrm{s}}}\log_2\left(\frac{p_{\mathrm{d}}}{p_{\mathrm{d}}+(1-p_{\mathrm{d}})p_{\mathrm{s}}}\right)\nonumber\\
    &\triangleq B_0.
    \label{eq:part1Entropy}
\end{align}
Similarly, for the summation over $w_k\in\M_{\mathrm{s}}$ in \eqref{eq:condEntropyGivenw2} we obtain
\begin{equation}
    p(W_k\in\mathcal{M}_{\mathrm{s}}\vert w_1^{k-1}\in A^{k-1}, T_{K}^{N}) = \frac{(1-p_{\mathrm{d}})p_{\mathrm{s}}}{p_{\mathrm{d}}+(1-p_{\mathrm{d}})p_{\mathrm{s}}} = 1-p(W_k\in\mathcal{M}_{\mathrm{a}}\vert w_1^{k-1}\in A^{k-1}, T_{K}^{N}).
    \label{eq:pWinMsGivenAT}
\end{equation}
Since the standard messages are not mutually exclusive, and equally likely, it follows that the second term in \eqref{eq:condEntropyGivenw2} becomes
 \begin{align}
 \begin{split}
    -\sum_{w_k\in\M_{\mathrm{s}}}p(w_k\vert w_1^{k-1},T_K^N)\log_2(p(w_k\vert w_1^{k-1},T_K^N)) &= -\sum_{w_k\in\M_{\mathrm{s}}}\frac{1}{M_{\mathrm{s}}}\frac{(1-p_{\mathrm{d}})p_{\mathrm{s}}}{p_{\mathrm{d}}+(1-p_{\mathrm{d}})p_{\mathrm{s}}}\log_2\left(\frac{1}{M_{\mathrm{s}}}\frac{(1-p_{\mathrm{d}})p_{\mathrm{s}}}{p_{\mathrm{d}}+(1-p_{\mathrm{d}})p_{\mathrm{s}}}\right)\\
    &= \frac{(1-p_{\mathrm{d}})p_{\mathrm{s}}}{p_{\mathrm{d}}+(1-p_{\mathrm{d}})p_{\mathrm{s}}}\left(\log_2M_{\mathrm{s}}-\log_2\left(\frac{(1-p_{\mathrm{d}})p_{\mathrm{s}}}{p_{\mathrm{d}}+(1-p_{\mathrm{d}})p_{\mathrm{s}}}\right)\right)\\
    &\triangleq B_1,
    \end{split}\label{eq:part2Entropy}
 \end{align}
Substituting \eqref{eq:part1Entropy} and \eqref{eq:part2Entropy} into \eqref{eq:condEntropyGivenw2} yields $H_{\mathrm{A}}(W_k\vert W_1^{k-1}=w_1^{k-1})=B_0+B_1$.

We now derive an expression for $p_{\mathrm{A}}(w_{1}^{k-1}\vert T_{K}^N)$ in \eqref{eq:generalCondEntSplit}. Let $1\le i\le k-1$ denote the (random) number of alarm messages in $w_1^{k-1}$ and, without loss of generality, assume that the alarm messages occupy the first $i$ positions of $W_1^{k-1}$, i.e. $w_1,\ldots,w_i\in[\mathcal{M}_{\mathrm{a}}]^i$ and $w_{i+1},\ldots,w_{k-1}\in[\mathcal{M}_{\mathrm{s}}]^{k-i+1}$. For a fixed $i$, the probability $p_{\mathrm{A}}(w_{1}^{k-1}\vert T_{K}^N)$ is obtained using Bayes' theorem as
\begin{align}
        &p_{\mathrm{A}}(W_1^i=w_1^{i}\in[\mathcal{M}_{\mathrm{a}}]^{i},W_{i+1}^{k-1} =w_{i+1}^{k-1}\in[\mathcal{M}_{\mathrm{s}}]^{k-(i+1)}\vert T_{K}^{N})\nonumber\\
        &= \frac{1}{M_{\mathrm{a}}M_{\mathrm{s}}^{k-(i+1)}}\frac{(p_{\mathrm{d}}+(1-p_{\mathrm{d}})p_{\mathrm{s}})^{K-(k-1)}(1-p_{\mathrm{d}})^{N-K}p_{\mathrm{a}}p_{\mathrm{d}}^{i}(1-p_{\mathrm{d}})^{k-(i+1)}p_{\mathrm{s}}^{k-(i+1)}}{p_{\mathrm{a}}(p_{\mathrm{d}}+(1-p_{\mathrm{d}})p_{\mathrm{s}})^{K}(1-p_{\mathrm{d}})^{N-K}+(1-p_{\mathrm{a}})p_{\mathrm{s}}^{K}}\nonumber\\
        &= \frac{p_{\mathrm{a}} p_{\mathrm{d}}^i((1-p_{\mathrm{d}})p_{\mathrm{s}})^{k-1-i}}{M_{\mathrm{a}}M_{\mathrm{s}}^{k-1-i}}N_0
    \label{eq:distributionUncond}
\end{align}
where $N_0$ is given as in \eqref{eq:N0}. Notice that as before only one alarm message is used at a given time so $M_{\mathrm{a}}$ is not raised to the power of $i$. Since there are exactly ${k-1\choose i}M_{\mathrm{a}}M_{\mathrm{s}}^{k-1-i}$ equiprobable and disjoint message sets $w_1^{k-1}$ consisting of $i$ alarm messages and $k-1-i$ standard messages, the first term of \eqref{eq:generalCondEntSplit} can be expressed as
\begin{align}
    \sum_{w_1^{k-1}\in A^{k-1}} p_{\mathrm{A}}(w_1^{k-1}\vert T_K^N)H_{\mathrm{A}}(W_k\vert W_1^{k-1}=w_1^{k-1}) &= \sum_{i=1}^{k-1}{k-1\choose i}\sum_{\substack{w_1^{i}\in [\M_{\mathrm{a}}]^i\\w_{i+1}^{k-1}\in[\M_{\mathrm{s}}]^{k-1-i}}} \frac{p_{\mathrm{a}} p_{\mathrm{d}}^i((1-p_{\mathrm{d}})p_{\mathrm{s}})^{k-1-i}}{M_{\mathrm{a}}M_{\mathrm{s}}^{k-1-i}}N_0(B_0+B_1)\\
    &= (B_0+B_1)\sum_{i=1}^{k-1}{k-1\choose i} p_{\mathrm{a}} p_{\mathrm{d}}^i((1-p_{\mathrm{d}})p_{\mathrm{s}})^{k-1-i}N_0.\label{eq:generalCondEntropyPart1}
\end{align}

We now consider the second term in \eqref{eq:generalCondEntSplit}. Here the conditional messages in $H_{\mathrm{S}}$ and messages in $p_{\mathrm{S}}$ are all standard messages. In contrast to the previous case, this can happen both when there is no alarm, and when there is alarm but none of the devices detect it. As before, we rewrite \eqref{eq:condEntropyGivenw} as
 \begin{align}
     \begin{split}
        H_{\mathrm{S}}(W_{k}\vert W_1^{k-1} = w_{1}^{k-1}) &= -\sum_{w_{k}\in\mathcal{M}_{\mathrm{a}}}p(w_{k}\vert w_1^{k-1}\in [\M_{\mathrm{s}}]^{k-1},T_{K}^{N})\log_2(p(w_{k}\vert w_1^{k-1}\in [\M_{\mathrm{s}}]^{k-1}, T_{K}^{N}))\\
        &\quad-\sum_{w_{k}\in\mathcal{M}_{\mathrm{s}}}p(w_{k}\vert w_1^{k-1}\in [\M_{\mathrm{s}}]^{k-1},T_{K}^{N})\log_2(p(w_{k}\vert w_1^{k-1}\in [\M_{\mathrm{s}}]^{k-1}, T_{K}^{N})).
        \label{eq:condEntropyGivenwS}
    \end{split}
\end{align}
Since each alarm message is equally likely, applying Bayes' theorem and the law of total probability repeatedly yields
\begin{align}
    p(w_k\in\mathcal{M}_{\mathrm{a}}\vert w_1^{k-1}\in[\mathcal{M}_{\mathrm{s}}]^{k-1}, T_{K}^{N})&=\frac{1}{M_{\mathrm{a}}} \frac{p_{\mathrm{a}} p_{\mathrm{d}}(p_{\mathrm{d}}+(1-p_{\mathrm{d}})p_{\mathrm{s}})^{K-k}(1-p_{\mathrm{d}})^{N-K+k-1}p_{\mathrm{s}}^{k-1}}{p_{\mathrm{a}}(p_{\mathrm{d}}+(1-p_{\mathrm{d}})p_{\mathrm{s}})^{K-k+1}(1-p_{\mathrm{d}})^{N-K+k-1}p_{\mathrm{s}}^{k-1}+(1-p_{\mathrm{a}})p_{\mathrm{s}}^K}\\
    &\triangleq \frac{1}{M_{\mathrm{a}}}B_3.
    \label{eq:pWinMaGivenWinMs}
\end{align}
Similarly, for $p(w_k\in\M_{\mathrm{s}}\vert w_1^{k-1}\in[\M_{\mathrm{s}}]^{k-1}, T_K^N)$ we obtain
\begin{equation}
    p(w_k\in\mathcal{M}_{\mathrm{s}}\vert w_1^{k-1}\in[\mathcal{M}_{\mathrm{s}}]^{k-1}, T_{K}^{N}) = \frac{1}{M_{\mathrm{s}}}(1-B_3).
   \label{eq:pWinMsGivenWinMs}
\end{equation}
Therefore we get 
\begin{align}
    H_{\mathrm{S}}(W_k\vert W_1^{k-1}=w_1^{k-1}) &= -\sum_{w_{k}\in\mathcal{M}_{\mathrm{a}}}\frac{B_3}{M_{\mathrm{a}}}\log_2\frac{B_3}{M_{\mathrm{a}}}-\sum_{w_{k}\in\mathcal{M}_{\mathrm{s}}}\frac{1-B_3}{M_{\mathrm{s}}}\log_2\frac{1-B_3}{M_{\mathrm{s}}}\\
    &= -B_3\log_2\frac{B_3}{M_{\mathrm{a}}}-(1-B_3)\log_2\frac{1-B_3}{M_{\mathrm{s}}}.
    \label{eq:condEntropyGivenwS2}
\end{align}
Finally, $p_{\mathrm{S}}(w_1^{k-1}\vert T_K^N)$ is given by
\begin{align}
    \begin{split}
        p(w_1^{k-1} \in[\mathcal{M}_{\mathrm{s}}]^{k-1}\vert T_{K}^N) &= \frac{1}{M_{\mathrm{s}}^{k-1}}\frac{p_{\mathrm{a}}(1-p_{\mathrm{d}})^{N-K+(k-1)}p_{\mathrm{s}}^{k-1}(p_{\mathrm{d}}+(1-p_{\mathrm{d}})p_{\mathrm{s}})^{K-(k-1)}+(1-p_{\mathrm{a}})p_{\mathrm{s}}^{K}}{p_{\mathrm{a}}(p_{\mathrm{d}}+(1-p_{\mathrm{d}})p_{\mathrm{s}})^{K}(1-p_{\mathrm{d}})^{N-K}+(1-p_{\mathrm{a}})p_{\mathrm{s}}^{K}}\\
    &\triangleq \frac{1}{M_{\mathrm{s}}^{k-1}}B_2.
    \end{split}
    \label{eq:probAllStandardMes}
\end{align}
Using \eqref{eq:condEntropyGivenwS2} and \eqref{eq:probAllStandardMes}, the last term in \eqref{eq:generalCondEntSplit} can be expressed as
\begin{align}
    \sum_{w_1^{k-1}\in[\M_{\mathrm{s}}]^{k-1}}p_{\mathrm{S}}(w_1^{k-1}\vert T_K^N)H_{\mathrm{S}}(W_k\vert W_1^{k-1}=w_1^{k-1}) &= \sum_{w_1^{k-1}\in[\M_{\mathrm{s}}]^{k-1}}\frac{B_2}{M_{\mathrm{s}}^{k-1}}\left(-B_3\log_2\frac{B_3}{M_{\mathrm{a}}}-(1-B_3)\log_2\frac{1-B_3}{M_{\mathrm{s}}}\right)\\
    &= -B_2\left(B_3\log_2\frac{B_3}{M_{\mathrm{a}}}+(1-B_3)\log_2\frac{1-B_3}{M_{\mathrm{s}}}\right)\label{eq:generalCondEntropyPart2}
\end{align}
Inserting \eqref{eq:generalCondEntropyPart1} and \eqref{eq:generalCondEntropyPart2} into \eqref{eq:generalCondEntropy} yields the final expression:
\begin{equation}
    H(W_k\vert W_1^{k-1}) =
     (B_0+B_1)\sum_{i=1}^{k-1}{k-1\choose i} p_{\mathrm{a}} p_{\mathrm{d}}^i((1-p_{\mathrm{d}})p_{\mathrm{s}})^{k-1-i}N_0
    -B_2\left(B_3\log_2\frac{B_3}{M_{\mathrm{a}}}+(1-B_3)\log_2\frac{1-B_3}{M_{\mathrm{s}}}\right)
\end{equation}
\qed

 \section{Proof of Theorem~\ref{theo:RandomCodingBound}}\label{app:proofRandomCodingBound}
Generate the $M_{\mathrm{a}} + M_{\mathrm{s}} = M$ codewords $\vec c_1, \ldots, \vec c_M\stackrel{i.i.d.}{\sim}\mathcal{N}(\vec 0, P'\vec I_{n})$. We assume that the first $M_{\mathrm{a}}$ codewords are alarm messages and the last $M_{\mathrm{s}}$ codewords are standards messages. Recall that $W_j$ is the codeword selected by the $j$-th device. If $\Vert\vec c_{W_j}\Vert_{2}^{2}>nP$ then device $j$ transmits $\vec 0$ instead, i.e. $\vec X_j = \vec 0$, otherwise $\vec X_j = \vec c_{W_j}$, $1\leq j\leq K$. Decoding is done in two steps. The transmitted alarm message (if any) is decoded first and canceled from the received signal, and then the decoder proceeds to decoding the standard messages. Notice that the order of decoding reflects the fact that alarm messages are expected to have a higher reliability requirement than standard messages.

In the first step, the decoder estimates the transmitted alarm message $\widehat W$ and the number of devices transmitting the alarm message $\widehat{K}_{\mathrm{a}}$:
\begin{align}
    \widehat{W}, \widehat{K}_{\mathrm{a}} = \argmin_{\substack{W\in \M_{\mathrm{a}}\\ 0\leq K_{\mathrm{a}}\leq K}}\Vert K_{\mathrm{a}} \vec c_{W} - \vec Y\Vert_2^2.
    \label{eq:decoderMinimization}
\end{align}
If $\widehat K_{\mathrm{a}}\ge 1$ the decoder outputs $\widehat W$ as the alarm message and zero otherwise. The estimated interference from the alarm messages is subtracted from the received signal in a successive interference cancellation fashion as $\vec Y_{\text{SIC}} = \vec Y - \widehat K_{\mathrm{a}} \vec c_{\widehat W}$. Next the decoder outputs the set of standard messages $\hat{ \mathcal{S}}\in[\mathcal{M}_{\mathrm{s}}]^{K-\hat K_{\mathrm{a}}}$ 
\begin{align}
\hat {\mathcal{S}}=\argmin_{\mathcal{S}\in[\mathcal{M}_{\mathrm{s}}]^{K-\hat K_{\mathrm{a}}}}\norm{\sum_{W\in \mathcal{S}}\vec c_W-\vec Y_{\text{SIC}}}_2.
\end{align}

We consider Gallager type bounds~\cite{Gallager1965}. 
Initially we ignore the power constraint and assume that the transmitted codewords are $\vec X_j =\vec  c_{W_j}$ instead of $\vec X_j=\vec c_{W_j}\mathds{1}\{\norm{\vec c_{W_j}}_2<nP\}$. Furthermore, we assume that $W_j$ are drawn without replacement from $\M_{\mathrm{s}}$. The contributions from these assumptions will be taken into account later. We consider each of the four bounds in the theorem separately.

\subsection{Alarm decoding error}\label{appsec:proofRandomCodingBoundPart1}
We start with the bound in \eqref{eq:alarmErrorBound}, i.e. the probability of not decoding an alarm message in the alarm event, denoted by $\P{E_{\mathrm{a}}\vert A}$. From symmetry we assume that devices $1, \ldots, K_{\mathrm{a}}$ are transmitting alarm message $1= W_1 = W_2 = \cdots = W_{K_{\mathrm{a}}}$. We want to bound $\P{\widehat{W}\neq 1}$. 
Define the interference as $\vec S = \sum_{m=K_{\mathrm{a}}+1}^{K}\vec X_m$, where we note the dependency on $K_{\mathrm{a}}$. We then have $\vec Y = K_{\mathrm{a}}\vec X_1 + \vec S + \vec Z$. Let $W'$ be a random index in $\mathcal{M}_{\mathrm{a}}\setminus 1$ and let $0\leq K_{\mathrm{a}}'\leq K$ be some integer. Then by definition of the decoder \eqref{eq:decoderMinimization} an error occurs if
\begin{equation}
    \Vert K_{\mathrm{a}}'\vec c_{W'} - (K_{\mathrm{a}}\vec X_1 + \vec S + \vec Z)\Vert_2^2 < \Vert K_{\mathrm{a}}\vec X_1 - (K_{\mathrm{a}}\vec X_1 + \vec S + \vec Z)\Vert_2^2,
\end{equation}
i.e. if the distance, in $L_2$-norm, from a multiple of a wrong codeword $\vec c_{W'}$ to the received signal $\vec Y$ is smaller than from the $K_{\mathrm{a}}$ true alarm transmissions $\vec X_1$. We therefore define the error event
\begin{equation}
    F_{\mathrm{a}}(W', K_{\mathrm{a}}') = \{\Vert K_{\mathrm{a}}\vec X_1-K_{\mathrm{a}}'\vec c_{W'} + \vec S+\vec Z\Vert_2^2 < \Vert\vec S+\vec Z\Vert_2^2\}
\end{equation}
We want to bound the probability of this event for all possible combinations of $W'$ and $K_{\mathrm{a}}'$, thus we define the collection of events  
\begin{equation}
    F_{\mathrm{a}}(K_{\mathrm{a}}') = \bigcup_{W'\in{\mathcal{M}_{\mathrm{a}}\setminus 1}} F_{\mathrm{a}}(W',K_{\mathrm{a}}')
    \label{eq:fUnion}
\end{equation}
and 
\begin{equation}
    F_{\mathrm{a}} = \bigcup_{0\leq K_{\mathrm{a}}'\leq K}F_{\mathrm{a}}(K_{\mathrm{a}}').
\end{equation}
Clearly, $\P{F_{\mathrm{a}}} = \P{\widehat{W}\neq 1} = \P{E_{\mathrm{a}}\vert A}$ since the decoder is designed to only output one alarm message, thereby eliminating the possibility of collision of alarm messages at the decoder.

We first use the fact that $\vec S$ is a sum of Gaussian random vectors and hence is also Gaussian, and obtain the bound
\begin{align}
    \P{F_{\mathrm{a}}(W', K_{\mathrm{a}}')\vert \vec X_1, K_{\mathrm{a}}, \vec S, \vec Z} &\leq e^{\lambda_{\mathrm{a}}\Vert\vec S+\vec Z\Vert_2^2}\E[\vec c_{W'}]{e^{-\lambda_{\mathrm{a}}\Vert K_{\mathrm{a}}\vec X_1 - K_{\mathrm{a}}'\vec c_{W'} + \vec S + \vec Z\Vert_2^2}}\label{eq:firstChernoffBound}\\
                                                   &= e^{\lambda_{\mathrm{a}}\Vert\vec S+\vec Z\Vert_2^2}\frac{e^{-\frac{\lambda_{\mathrm{a}}\Vert K_{\mathrm{a}}\vec X_1 + \vec S+ \vec Z\Vert_2^2}{1+2K_{\mathrm{a}}'^{2}P'\lambda_{\mathrm{a}}}}}{(1+2K_{\mathrm{a}}'^{2}P'\lambda_{\mathrm{a}})^{n/2}}\label{eq:firstChernoffIdentity}\\
                                                   &= e^{\lambda_{\mathrm{a}}\Vert \vec S+\vec Z\Vert_2^2}e^{-\frac{\lambda_{\mathrm{a}}\Vert K_{\mathrm{a}}\vec X_1 + \vec S+ \vec Z\Vert_2^2}{1+2K_{\mathrm{a}}'^2P'\lambda_{\mathrm{a}}}}e^{-\frac{n}{2}\ln(1+2K_a'^2P'\lambda_{\mathrm{a}})}\\
                            &= e^{\lambda_{\mathrm{a}}\norm{\vec S+\vec Z}_2-\beta_{\mathrm{a}}\norm{K_a\vec X_1 +\vec S+\vec Z}_2-n\phi(K_a'^2,\lambda_{\mathrm{a}})},
   \label{eq:firstBound}
\end{align}
where $\lambda_{\mathrm{a}}\in\mathbb{R}_+$, $\phi(k, \alpha) = \frac{1}{2}\ln(1+2kP'\alpha)$ and $\beta_{\mathrm{a}}=\Phi(K_a'^2, \lambda_{\mathrm{a}})$ where $\Phi(k,\alpha)\triangleq \frac{\alpha}{1+2kP'\alpha}$. The bound in \eqref{eq:firstChernoffBound} follows from the Chernoff bound~\cite{Goemans2015} and \eqref{eq:firstChernoffIdentity} uses the identity \cite{Polyanskiy2017}
\begin{equation}
    \E{e^{-\gamma\Vert \sqrt{\alpha}\vec Q + \vec u\Vert_2^2}} = \frac{e^{-\frac{\gamma\Vert\vec u\Vert_2^2}{1+2\alpha\gamma}}}{(1+2\alpha\gamma)^{n/2}},
   \label{eq:chernoffIdentity}
\end{equation}
where $\vec u\in\mathbb{R}^{n}$, $\alpha\in\mathbb{R}_+$, $\gamma\geq-\frac{1}{2\alpha}$ and $\vec Q\sim\mathcal{N}(\vec 0, \vec I_n$).

Next we use Gallager's $\rho$-trick~\cite{Gallager1965} to bound $\P{F_{\mathrm{a}}(K_{\mathrm{a}}')}$. For events $A_1,A_2,\ldots$ and $\rho\in[0,1]$ we have $\P{\cup_jA_j}\leq\left(\sum_{j}\P{A_j}\right)^{\rho}$. We get
\begin{equation}
    \P{F_{\mathrm{a}}(K_{\mathrm{a}}')\vert \vec X_1, K_{\mathrm{a}}, \vec S, \vec Z}\leq \left(M_{\mathrm{a}}-1\right)^{\rho}e^{\rho\lambda_{\mathrm{a}}\Vert \vec S+\vec Z\Vert_2^2-\rho\beta_{\mathrm{a}}\Vert K_{\mathrm{a}}\vec X_1 + \vec S+ \vec Z\Vert_2^2 - \rho n\phi(K_{\mathrm{a}}'^2, \lambda_{\mathrm{a}})}.
\end{equation}
Taking expectation over $\vec X_1$ and using \eqref{eq:chernoffIdentity} yields
\begin{align}
    \P{F_{\mathrm{a}}(K_{\mathrm{a}}')\vert K_{\mathrm{a}}, \vec S, \vec Z} &\leq (M_{\mathrm{a}}-1)^{\rho}e^{\rho\lambda_{\mathrm{a}}\norm{\vec S+\vec Z}_2}\E[\vec X_1]{e^{-\rho\beta_{\mathrm{a}}\norm{K_{\mathrm{a}}\vec X_1+\vec S+\vec Z}_2}}e^{-\rho n\phi(K_{\mathrm{a}}'^2,\lambda_{\mathrm{a}})}\label{eq:1expectOverX1}\\
    &= (M_{\mathrm{a}}-1)^{\rho}e^{\rho\lambda_{\mathrm{a}}\norm{\vec S+ \vec Z}_2}\frac{e^{-\frac{\rho\beta_{\mathrm{a}}\norm{\vec S+\vec Z}_2}{1+2K_a^2P'\rho\beta_{\mathrm{a}}}}}{(1+2K_a^2P'\rho\beta_{\mathrm{a}})^{n/2}}e^{-\rho n\phi(K_a'^2,\lambda_{\mathrm{a}})}\\
    &= (M_{\mathrm{a}}-1)^{\rho}e^{-\gamma_{\mathrm{a}}\Vert \vec S+\vec Z\Vert_2^2-n\tau}\label{eq:3expectOverX1},
\end{align}
where $\tau = \rho \phi(K_{\mathrm{a}}'^2, \lambda_{\mathrm{a}}) + \phi(K_{\mathrm{a}}^2, \rho\beta_{\mathrm{a}})$ and $\gamma_{\mathrm{a}}=\Phi(K_{\mathrm{a}}^2,\rho\beta_{\mathrm{a}})-\rho\lambda_{\mathrm{a}}$. Now in the same manner as in \eqref{eq:1expectOverX1}-\eqref{eq:3expectOverX1} expectation is taken over $\vec S$ and $\vec Z$ where \eqref{eq:chernoffIdentity} is used for both. We get
\begin{equation}
    \P{F_{\mathrm{a}}(K_{\mathrm{a}}')\vert K_{\mathrm{a}}} \leq e^{\rho\ln(M_{\mathrm{a}}-1) - n\xi_{\mathrm{a}}},
   \label{eq:FixedKaErrorBound}
\end{equation}
where $\xi_{\mathrm{a}}=\tau + \phi(K-K_{\mathrm{a}}, \gamma_{\mathrm{a}}) + \phi(1/P',\psi_{\mathrm{a}})$ and $\psi_{\mathrm{a}}=\Phi(K-K_{\mathrm{a}},\gamma_{\mathrm{a}})$. Introducing $E_{\mathrm{a}} = \max_{0\leq \rho\leq 1, 0<\lambda_{\mathrm{a}}}-\frac{\rho}{n}\ln(M_{\mathrm{a}}-1)+\xi_{\mathrm{a}}$ and applying the union bound gives
\begin{align}
    \P{F_{\mathrm{a}}\vert K_{\mathrm{a}}} &= \min\left(\sum_{K_{\mathrm{a}}'=0}^{K}e^{-nE_{\mathrm{a}}}, ~1\right)\\
    &\triangleq a(K,K_{\mathrm{a}}).\label{eq:a_proof}
\end{align}
Finally, we take the expectation over $K_{\mathrm{a}}$. The distribution of $K_{\mathrm{a}}$ is a binomial distribution given by
\begin{equation}
    p_{K_{\mathrm{a}}}(k) = {K \choose k} \frac{p_{\mathrm{d}}^k\left(\left(1-p_{\mathrm{d}}\right)p_{\mathrm{s}}\right)^{K-k}}{(p_{\mathrm{d}} + (1-p_{\mathrm{d}})p_{\mathrm{s}})^{K}},
\end{equation}
where the normalization coefficient arises because of the certainty that $K$ devices were active. It follows that
\begin{equation}
    \P{F_{\mathrm{a}}}\leq \sum_{K_{\mathrm{a}}=0}^{K}p_{K_{\mathrm{a}}}(K_{\mathrm{a}})a(K,K_{\mathrm{a}}).
\end{equation}

We now consider the impact of the power constraint. Since the standard messages are treated as interference in this bound, we ignore the power constraint for the standard messages as the bound is still valid. For the alarm messages only one is active at a given time, so we add the following term to the error probability:
\begin{align}
    \P{\Vert\vec c_j\Vert_2^2>nP}&= \P{\frac{1}{n}\sum_{i=1}^{n}Z_i^2 > \frac{P}{P'}}\\ &\triangleq p_0,
   \label{eq:totalVarianceDist}
\end{align}
where $\vec Z = [Z_1, \ldots, Z_n]^T\sim \mathcal{N}(\vec 0, \vec I_n)$. This gives the bound in \eqref{eq:alarmErrorBound}.

\subsection{False positive alarms}\label{appsec:proofRandomCodingBoundPart2}
We now turn to the bound in \eqref{eq:fPErrorBound}, i.e. the bound for the probability of false positive alarms, $\P{E_{\mathrm{fp}}\vert\neg A}$. In this case the true $K_{\mathrm{a}} = 0$, and a false positive occurs if the decoder outputs $\widehat K_{\mathrm{a}}>0$. Let $W'\in\M_{\mathrm{a}}$ and $0< K_{\mathrm{a}}'\leq K$. We define the error event
\begin{equation}
    F_{\mathrm{fp}}(W', K_{\mathrm{a}}') = \{\norm{\vec S-K_{\mathrm{a}}'\vec c_{W'}+\vec Z}_2 < \norm{\vec S + \vec Z}_2\}.
   \label{eq:FFp}
\end{equation}
The only difference between the error event $F_{\mathrm{fp}}(W', K_{\mathrm{a}}')$ and $F_{\mathrm{a}}(W', K_{\mathrm{a}}')$ is the absence of the true alarm messages. Therefore we define
\begin{equation}
    F_{\mathrm{fp}}(K_{\mathrm{a}}') = \bigcup_{W'\in\M_{\mathrm{a}}} F_{\mathrm{fp}}(W', K_{\mathrm{a}}')
\end{equation}
and
\begin{equation}
    F_{\mathrm{fp}} = \bigcup_{0< K_{\mathrm{a}}'\leq K} F_{\mathrm{fp}}(K_{\mathrm{a}}').
\end{equation}
We have that $\P{F_{\mathrm{fp}}} = \P{E_{\mathrm{fp}}\vert \neg A}$. As in Section~\ref{appsec:proofRandomCodingBoundPart1} we use the Chernoff bound and the identity \eqref{eq:chernoffIdentity}, and take the expectation over $\vec c_{W'}$ to get the bound
\begin{align}
    \P{F_{\mathrm{fp}}(W', K_{\mathrm{a}}')\vert\vec S, \vec Z} &\leq e^{\lambda_{\mathrm{fp}}\norm{\vec S+\vec Z}_2}e^{\frac{-\lambda_{\mathrm{fp}}\norm{\vec S+\vec Z}_2}{1+2K_{\mathrm{a}}'^2P'\lambda_{\mathrm{fp}}}-n\phi(K_{\mathrm{a}}'^2,\lambda_{\mathrm{fp}})}\\
    &= e^{-\beta_{\mathrm{fp}}\norm{\vec S+\vec Z}_2 - n\phi(K_{\mathrm{a}}'^2, \lambda_{\mathrm{fp}})},
\end{align}
where $\lambda_{\mathrm{fp}}\in\mathbb{R}_+$ and $\beta_{\mathrm{fp}}=\Phi(K_{\mathrm{a}}'^2,\lambda_{\mathrm{fp}})-\lambda_{\mathrm{fp}}$. Using Gallager's $\rho$-trick we obtain
\begin{equation}
    \P{F_{\mathrm{fp}}(K_{\mathrm{a}}')\vert\vec S, \vec Z} \leq e^{\rho\ln(M_{\mathrm{a}})-\rho\beta_{\mathrm{fp}}\norm{\vec S+\vec Z}_2 - \rho n\phi(K_{\mathrm{a}}'^{2}, \lambda_{\mathrm{fp}})}.
\end{equation}
Taking the expectation over $\vec S$ and $\vec Z$, and applying the union bound over $K_{\mathrm{a}}'$, we get the bound 
\begin{align}
    \P{F_{\mathrm{fp}}}&\leq \min\left(\sum_{K_{\mathrm{a}}'=1}^{K}e^{-nE_{\mathrm{fp}}},~1\right)\\
    \triangleq& b(K)\label{eq:FPbound},
\end{align}
where $E_{\mathrm{fp}} = \max_{0\leq\rho\leq t, 0<\lambda_{\mathrm{fp}}}-\frac{\rho}{n}\ln M_{\mathrm{a}}+\xi_{\mathrm{fp}}$ for $\xi_{\mathrm{fp}}
 = \rho \phi(K_{\mathrm{a}}'^2, \lambda_{\mathrm{fp}}) + \phi(K, \rho\beta_{\mathrm{fp}}) + \phi(1/P', \gamma_{\mathrm{fp}})$ and $\gamma_{\mathrm{fp}}=\Phi(K,\rho\beta_{\mathrm{fp}})$. 

Similar to the previous case, constraining the transmission power results in less interference and hence lower error probability. Therefore, the bound given by \eqref{eq:FPbound} is still an upper bound on the error probability in the power constrained case, and no additional term is needed.

\subsection{Standard message error with no alarm}\label{appsec:proofRandomCodingBoundPart3}
We now consider the bound in \eqref{eq:noAlarmErrorBound}, i.e. the bound for the per-device probability of error for standard messages when no alarm has occurred, $\frac{1}{K}\sum_{j=1}^{K}\P{E_j\vert\neg A}$. Since the standard message decoder relies on canceling the interference caused by the alarm messages, we assume that correct decoding of standard messages can only occur if the decoder does not output a false positive alarm. If there is no false positive, the scenario coincides with the one derived in~\cite{Polyanskiy2017} given by $c(K)$ in  \eqref{eq:yuriErrorBound}. Since the probability of a false positive alarm is bounded by $b(K)$, the probability of error is bounded as 
\begin{align}
   \frac{1}{K}\sum_{j=1}^{K}\P{E_j\vert\neg A}&\leq 1-(1-b(K))(1-c(K))\\
   &= b(K)+c(K)-b(K)c(K))
\end{align}

The bound above ignores the impact of power constraint for the interfering standard messages. However, as in the previous section, the bound is still valid in the case of a false positive alarm. When no false positive alarm is decoded, the power constraint and collision error of standard messages are accounted for in $c(K)$ through the last two terms in \eqref{eq:yuriErrorBound} as derived in~\cite{Polyanskiy2017}.

\subsection{Standard message error with alarm}\label{appsec:proofRandomCodingBoundPart4}
Finally, we turn to the bound in \eqref{eq:AlarmStdErrorBound}, i.e. the average per-device probability of error for the standard messages in the alarm event, $\E[K_{\mathrm{a}}]{\frac{1}{K-K_{\mathrm{a}}}\sum_{j=1}^{K-K_{\mathrm{a}}}\P{E_j\vert A}}$. As in Section~\ref{appsec:proofRandomCodingBoundPart3}, we assume that there is automatically an error if the alarm is incorrectly decoded. Assume that the first $K_{\mathrm{a}}$ devices are transmitting the alarm message $1 = W_1 = \cdots = W_{K_{\mathrm{a}}}$. The probability of error for standard messages is bounded by the probability that the alarm is incorrectly decoded or the standard messages are incorrectly decoded after correctly canceling the interference from alarm messages. Since the probability for the latter event is bounded by $c(K-K_{\mathrm{a}})$ we obtain
\begin{align}
    \frac{1}{K-K_{\mathrm{a}}}\sum_{j=1}^{K-K_{\mathrm{a}}}\P{E_j\vert A}&\le 1-\P{\widehat W=1, \widehat K_{\mathrm{a}} = K_{\mathrm{a}}}(1-c(K-K_{\mathrm{a}}))\\
    &=1-\P{\widehat K_{\mathrm{a}}= K_{\mathrm{a}}\vert \widehat W=1}\P{\widehat W=1}(1-c(K-K_{\mathrm{a}}))\\
    &\le 1-\P{\widehat K_{\mathrm{a}}= K_{\mathrm{a}}\vert \widehat W=1}(1-a(K, K_{\mathrm{a}}))(1-c(K-K_{\mathrm{a}})),
   \label{eq:dBound}
\end{align}
where $a(K,K_{\mathrm{a}})$ is given by \eqref{eq:a_proof}. 

To derive a bound on $\P{\widehat K_{\mathrm{a}} = K_{\mathrm{a}}\vert \widehat W=1}$ we consider the complementary event $\P{\widehat K_{\mathrm{a}} \neq K_{\mathrm{a}}\vert \widehat W=1}$. Let $0\leq K_{\mathrm{a}}'\leq K$ and $K_{\mathrm{a}}'\neq K_{\mathrm{a}}$, and define error event 
\begin{equation}
    F_{\mathrm{sa}}(K_{\mathrm{a}}') = \{\norm{(K_{\mathrm{a}}-K_{\mathrm{a}}')\vec X_1 + \vec S + \vec Z}_2<\norm{\vec S + \vec Z}_2 \}.
\end{equation}
This event is similar to the error event $F_{\mathrm{a}}(W', K_{\mathrm{a}}')$ with the exception that here the alarm message is known. We define
\begin{equation}
    F_{\mathrm{sa}} = \bigcup_{\substack{0\leq K_{\mathrm{a}}'\leq K\\K_{a}'\neq K_{\mathrm{a}}}} F_{\mathrm{sa}}(K_{\mathrm{a}}'),
\end{equation}
and, as in Section~\ref{appsec:proofRandomCodingBoundPart1}, use the Chernoff bound and the identity \eqref{eq:chernoffIdentity} with the expectation over $\vec X_1$ to get the bound
\begin{align}
    \P{F_{\mathrm{sa}}(K_{\mathrm{a}}')\vert \vec S, \vec Z, K_{\mathrm{a}}} &\leq e^{\lambda_{\mathrm{sa}}\norm{\vec S+\vec Z}_2}e^{\frac{-\lambda_{\mathrm{sa}}\norm{\vec S+\vec Z}_2}{1+2(K_{\mathrm{a}}-K_{\mathrm{a}}')^2P'\lambda_{\mathrm{sa}}}}- e^{-\frac{n}{2}\ln(1+2(K_{\mathrm{a}}-K_{\mathrm{a}}')^2P'\lambda_{\mathrm{sa}}}\\
    &= -e^{-\beta_{\mathrm{sa}}\norm{\vec S+\vec Z}_2 - n\phi((K_{\mathrm{a}}-K_{\mathrm{a}}')^2, \lambda_{\mathrm{sa}})},
\end{align}
where $\beta_{\mathrm{sa}} = \Phi\left((K_{\mathrm{a}}-K_{\mathrm{a}}')^2 ,\lambda_{\mathrm{sa}}\right)-\lambda_{\mathrm{sa}}$. Taking the expectation over $\vec S$ and $\vec Z$ using \eqref{eq:chernoffIdentity} we have
\begin{equation}
    \P{F_{\mathrm{sa}}(K_{\mathrm{a}}')\vert K_{\mathrm{a}}} \leq e^{-nE_{\mathrm{sa}}},
\end{equation}
where $E_{\mathrm{sa}} = \max_{0<\lambda_{\mathrm{sa}}}\phi\left((K_{\mathrm{a}}-K_{\mathrm{a}}')^2,\lambda_{\mathrm{sa}}\right) + \phi(K-K_{\mathrm{a}},\beta_{\mathrm{sa}})+\phi(1/P',\gamma_{\mathrm{sa}})$ and $\gamma_{\mathrm{sa}} = \Phi(K-K_{\mathrm{a}},\beta_{\mathrm{sa}})$. Finally, the union bound over $K_{\mathrm{a}}'$ is used to get 
\begin{align}
    \P{F_{\mathrm{sa}}\vert K_{\mathrm{a}}}&\leq \min\left(\sum_{\substack{K_{\mathrm{a}}'=0\\K_{\mathrm{a}}'\neq K_{\mathrm{a}}}}^{K}e^{-nE_{\mathrm{sa}}},~1\right)\\
    &\triangleq e(K,K_{\mathrm{a}}).
\end{align}
It follows that $\P{\widehat K_{\mathrm{a}}=K_{\mathrm{a}}\vert \widehat W=1}\geq
1-e(K,K_{\mathrm{a}})$. Substituting this into \eqref{eq:dBound} and taking expectation over $K_{\mathrm{a}}$ as in Section~\ref{appsec:proofRandomCodingBoundPart1} gives \begin{equation}
    \E[K_{\mathrm{a}}]{\frac{1}{K-K_{\mathrm{a}}}\sum_{j=1}^{K-K_{\mathrm{a}}}\P{E_j\vert A}}\leq \sum_{K_{\mathrm{a}}=0}^{K}p_{K_{\mathrm{a}}}\left(K_{\mathrm{a}}\right)\left(1-(1-a(K,K_{\mathrm{a}}))(1-e(K,K_{\mathrm{a}}))\left(1-c\left(K-K_{\mathrm{a}}\right)\right)\right).
\end{equation}

The bound above already includes collisions and the power constraint in the decoding of standard messages through $c\left(K-K_{\mathrm{a}}\right)$. However, we still need to include the power constraint of the alarm messages. As in Section~\ref{appsec:proofRandomCodingBoundPart1}, this is done by adding $p_0$ to each of the two error event bounds $e(K,K_{\mathrm{a}})$ and $a(K,K_{\mathrm{a}})$. By defining $d(K,K_{\mathrm{a}})\triangleq (1-(a(K,K_{\mathrm{a}})+p_0))(1-(e(K,K_{\mathrm{a}})+p_0))$ the final bound becomes
\begin{equation}
    \E[K_{\mathrm{a}}]{\frac{1}{K-K_{\mathrm{a}}}\sum_{j=1}^{K-K_{\mathrm{a}}}\P{E_j\vert A}}\leq \sum_{K_{\mathrm{a}}=0}^{K}p_{K_{\mathrm{a}}}\left(K_{\mathrm{a}}\right)\left(1-d\left(K,K_{\mathrm{a}}\right)\left(1-c\left(K-K_{\mathrm{a}}\right)\right)\right).
\end{equation}
\qed 
\section*{Acknowledgment}
This work has been in part supported the European Research Council (ERC) under the European Union Horizon 2020 research and innovation program (ERC Consolidator Grant Nr. 648382 WILLOW) and Danish Council for Independent Research (Grant Nr. 8022-00284B SEMIOTIC).

\ifCLASSOPTIONcaptionsoff
  \newpage
\fi

\end{document}